\documentclass[final,authoryear,5p,times,twocolumn]{elsarticle}

\usepackage{amssymb}
\usepackage{amsmath}
\usepackage{hyperref} 
\hypersetup{colorlinks,citecolor=blue,linkcolor=blue,urlcolor=blue}

\journal{Astronomy and Computing}

\def\fig{Figure}
\def\figs{Figures.}
\def\Fig{Figure}

\def\sects{Sections}

\def\tab{Table}

\def\eq{eq}

\def\aegean{{\sc Aegean}}
\def\bane{{\sc BANE}}
\def\mimas{{\sc MIMAS}}
\def\aeres{{\sc AeRes}}
\def\fitswarp{{\sc Fits\_Warp}}
\def\tool{{\sc Robbie}}

\def\github{{https://github.com/PaulHancock/}}

\begin{document}

\begin{frontmatter}



%\title{BAtch procesSing for Transient And vaRiable Detection: \tool{}}
\title{\tool: A Batch Processing Work-flow for the Detection of Radio Transients and Variables}


% \author[Place]{I. M. First}
% \ead{first@mail.com}
% \author[Place]{I. R. Second}
% \address[Place]{The University of Learned Folks}

\author[ICRAR]{P. J. Hancock}
\ead{Paul.Hancock@curtin.edu.au} % orcid 0000-0002-4203-2946
\author[ICRAR]{N. Hurley-Walker} % orcid
\author[ICRAR]{T. E. White}
\address[ICRAR]{International Center for Radio Astronomy Research - Curtin University}

\begin{abstract}
We present \tool{}: a general work-flow for the detection and characterization of radio variability and transient events in the image domain.
\tool{} is designed to work in a batch processing paradigm with a modular design so that components can be swapped out or upgraded to adapt to different input data, whilst retaining a consistent and coherent methodological approach.
\tool{} is based on commonly used and open software, and is encapsulated in a Makefile to aid portability and reproducibility.
In this paper we describe the methodology behind \tool{}, and demonstrate its use on real and simulated data.
\tool{} is available on \href{\github{}Robbie}{GitHub}.

\end{abstract}
 

\end{frontmatter}

% \linenumbers

\section{Introduction}
\label{intro}

Much effort has been put into the development of real-time pipelines for the detection of variable and transient radio events in the image domain \citep[eg.][]{banyer_vast_2012,Swinbank_lofar_2015}.
However, since the input images cannot yet be produced in real time, these stream-based processing approaches are often run offline, multiple times, with results only being reported once all the data has been captured.
Thus, in the absence of real-time calibration and imaging, a real-time / stream based approach to detection and characterization is not essential and a robust batch processing approach is sufficient.
By moving to a batch processing paradigm the work-flow is simpler to conceive and implement as the salient properties (number of epochs, number of dimensions, dynamic range, noise properties, etc.) of the data is known in advance.

Blind surveys for variable and transient radio sources have typically fallen into two categories: opportunistic surveys conducted with data observed for other science projects \citep[e.g. ][]{bannister_22yr_2011,hancock_radio_2016}; and deliberate surveys for which the observations were planned to target a particular transient or variable source population \citep[e.g. ][]{Bell_survey_2014,rowlinson_limits_2016}.

One of the difficulties in the analysis of surveys for radio variability and transients is the use of differing statistical methodologies between the various works.
Early work focused on the modulation index ($m=\frac{\sigma}{\mu}$), which was used both as an indication of significance and magnitude of variability, and used a cut-off to denote a fraction of sources as variable.
The literature has evolved somewhat to use some more statistically useful criteria for the detection of variables, such as: a de-biased modulation index; quoting an areal density of variable and transient rates; and to begin to decouple the incidence and significance of variability.
The phase space that has been explored by variable and transient surveys has also expanded to differentiate between surveys with different sensitivities, frequencies, and cadences \citep[see e.g.][]{rowlinson_limits_2016, mooley_survey_2016}.
As the standardization of statistical methodology continues, it becomes easier to compare the various radio surveys.
Whilst the over-arching data processing methodology employed by radio variability surveys is quite similar, there has yet to be a standard technique employed, and indeed no reproducible methodology has been published.
None of the studies referenced by \citet{mooley_survey_2016}\footnote{For an updated list see: www.tauceti.caltech.edu/kunal/radio-transient-surveys/} correspond to a reproducible scientific result: the data, code, and methodology have not been provided in a manner that can be replicated exactly.
Data and software requirements are often provided and the methodology is described \citep[as in ][]{hancock_radio_2016}, however this does not make a reproducible result.

In order to improve the reproducibility of future radio variability surveys, and to provide a more detailed description of many of the aspects that have been included in previous surveys, we present here a work-flow that can be adapted to a variety of surveys and instruments.
\tool{} has been explicitly designed to be simple to install and operate, and  easily extensible to new instruments.
% The advantage of a work-flow like \tool{} over a system like the VAST or TraP pipelines, is that the effort required to install, operate, and adapt the system to new instruments or data sets is significantly reduced.
In this regard \tool{} could be considered an entry-level processing pipeline, with many of the advantages of the more complex alternatives.

With \tool{} we provide a first step to making variable and transient results reproducible by encoding the methodology into a work-flow that can be easily shared and replicated. 
The work-flow embedded in \tool{} is currently being used to search for transients and variables in the Kepler K2 fields \citep[and Tingay et al. in prep]{Tingay_multi-resolution_2016}, and in a survey for interstellar scintillation (Hancock et al., in prep).
Both of these works rely on low-frequency observations with the Murchison Widefield Array \cite[MWA;][]{tingay_murchison_2013}, for which ionospheric disturbances cause a warping of the sky as seen by the telescope.
For this reason the work-flow includes an essential pre-processing stage that corrects for direction-dependent ionospheric distortions in the images.

We opt to use a Makefile\footnote{Following the \href{https://www.gnu.org/software/make/}{GNUMake} specification} as a work-flow manager.
{\sc Make} was designed to facilitate the compilation of C code \citep{feldman_make_1979} and will implement a given set of rules in order to create output targets, and will recreate intermediate files only as necessary.
The preservation of intermediate files is particularly useful as the work-flow can be restarted part-way though, allowing for a faster and more flexible development cycle.
However, our implementation is outside the intended use case of {\sc Make}, which can limit or complicate the process of automating the work-flow.
Additionally, {\sc Make} is not easily scalable to a diversity of computing environments, is not able to scale across multiple nodes of an HPC, and does not interact with job scheduling software commonly used in HPC environments.
Furthermore, software versions and environments need to be managed in order for methodology to be truly reproducible.
\tool{} does not in and of itself preserve its environment, but the use of Docker\footnote{\href{www.docker.com}{http://www.docker.com}} containers has recently become a popular method by which to address this reproducibility problem.
The tasks carried out by \tool{} are not themselves dependent on the use of {\sc Make}, allowing \tool{} to be upgraded to a work-flow manager such as the Common Workflow Language \citep[CWL,][]{amustutz_common_2016}, which would address each of the previously mentioned issues, including the use of Docker containers and HPC job schedulers.

\section{Test Data}
\label{sec:data}
In \sects{}\,\ref{sec:methods}-\ref{sec:validation} we will demonstrate and validate the functionality of \tool{} using public data from the archive\footnote{\href{asvo.mwatelescope.org}{http://asvo.mwatelescope.org}} of the Murchison Widefield Array \citep[MWA,][]{tingay_murchison_2013}, and data which we have simulated for this project.
Here we describe these data.

\subsection{Observational Data}
\label{sec:obsdata}
\begin{figure}[ht]
\centering
\includegraphics[width=0.9\linewidth]{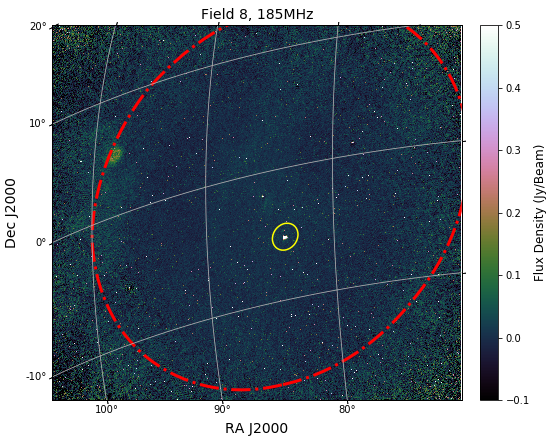}
\caption{
An image of the field-of-view and region of interest for the observational data.
The image is the mean of 25 epochs.
The red circle and the upper/right image boundaries define the region over which sources will be extracted, and transients discovered.
The Orion nebula is within this field and is indicated by the yellow circle.
}
\label{fig:skyplot}
\end{figure}

The observational test data consist of images collected at $185$\,MHz as part of a larger project to study interstellar scintillation in the Milky Way (MWA Project ID G0003, PI Hancock).
The region of interest is indicated in Figure\,\ref{fig:skyplot}.
Observations were conducted on a roughly weekly cadence, with observations being conducted at the same local sidereal time (LST) to ensure a consistent $(u,v)$ sampling.
The data were calibrated and flagged using an MWA processing pipeline derived from the GLEAM survey \citep{HurleyWalker_GLEAM_2017}, which included the use of {\sc WSClean} \citep{offringa_wsclean_2014}, and {\sc AOflagger} \citep{offringa_low-frequency_2015}.
The consistent LST of the observations means that the resulting images should have a consistent pixel to sky coordinate mapping.
In total 33 observations were conducted, but only 25 of these produced images that were of good enough quality to include in this work.
The observations that contributed to this work are listed in Table\,\ref{tab:observations}, and the corresponding raw data can be accessed via the MWA archive using the observation identifier (OBSID).

\begin{table}
\centering
\begin{tabular}{cc | cc}
\hline
OBSID & Date & OBSID & Date\\
      & 2013 &       & 2014 \\
\hline
1061674824 & Aug-27 & 1091401456 & Aug-06\\
1062277968 & Sep-03 & 1091918440 & Aug-12\\
1062881120 & Sep-10 & 1092521592 & Aug-19\\
1063484264 & Sep-17 & 1093038576 & Aug-25\\
1064690568 & Oct-01 & 1093727888 & Sep-02\\
1065293712 & Oct-08 & 1094331032 & Sep-09\\
1067103160 & Oct-29 & 1094761856 & Sep-14\\
1068481784 & Nov-14 & 1095537336 & Sep-23\\
1068912608 & Nov-19 & 1096140480 & Sep-30\\
1069515752 & Nov-26 & 1097346776 & Oct-14\\
1070118904 & Dec-03 & 1098208416 & Oct-24\\
           &        & 1099070064 & Nov-03\\
           &        & 1099759376 & Nov-11\\
           &        & 1100276360 & Nov-17\\
\hline
\end{tabular}
\caption{
A listing of the 25 MWA observations that were imaged and used as test data for this work.
The field was in the day time sky at the given LST during Jan-Jul of each year and so observations were not taken during these months.
}
\label{tab:observations}
\end{table}

\tool{} relies on \aegean{} \citep{Hancock_compact_2012,Hancock_source_2018} and \fitswarp{} \citep{hurley-walker_fitswarp_2018} for source-finding and image warping.
These tools in turn require the image data to be presented as FITS format images with world coordinate system (WCS) described in the header.
For catalogue handling, \aegean{} can read and write a multitude of table formats, but the FITS binary table format is preferred due to the smaller file size and reduced read/write times.
Finally, the source-finding performed by \aegean{} can be constrained to a sub-region of an image using a region file in a custom format produced by MIMAS\footnote{Part of the \href{https://github.com/PaulHancock/Aegean}{AegeanTools} library} \citep{Hancock_source_2018}.
Thus our observational test data consist of 25 images in FITS format, with WCS headers describing a SIN projection, as well as a region file that describes a circle of radius $15^\circ$ centered at $\mathrm{RA}=85^\circ$, $\mathrm{Dec}=-1^\circ$.
These data can be downloaded from \citep{Hancock_data_robbie_2019}.

\subsection{Simulated Data}
\label{sec:simdata}
The most direct validation of the abilities of \tool{} is to compare the extracted light curves to some {\it a-priori} known version of the 'true' light curves.
To achieve this we simulated a set of 25 observations with realistic properties, and a known set of variable, non-variable, and transient sources.
The simulation code is available on GitHub\footnote{\github\href{https://gitub.com/PaulHancock/SIMRobbie}{SIMRobbie}} and the simulated catalogues and images are available from \cite{Hancock_data_robbie_2019}.

The simulation proceeds as follows:
Create a square region of interest using MIMAS, centered at $(\mathrm{RA},\mathrm{Dec}) = (180^\circ,0^\circ)$, of width and height $7^\circ$.
Within the region of interest create a uniform distribution of 1100 positions.
Create source populations with flux densities logarithmically spaced between $5$\,mJy and $1$\,Jy consisting of 500 variable sources, 500 non-variable sources, and 100 transient sources.
For each source generate a light curve over 25 epochs that is either: variable with a modulation index of 5\%; non-variable with the same flux density in all epochs; or transient, with all epochs having zero flux density and a single epoch of non-zero flux density.
For each epoch generate an image that is slightly larger than the region of interest, which has a rms noise of $5$\,mJy\,beam$^{-1}$.

\begin{figure}[ht]
\centering
\includegraphics[width=0.9\linewidth]{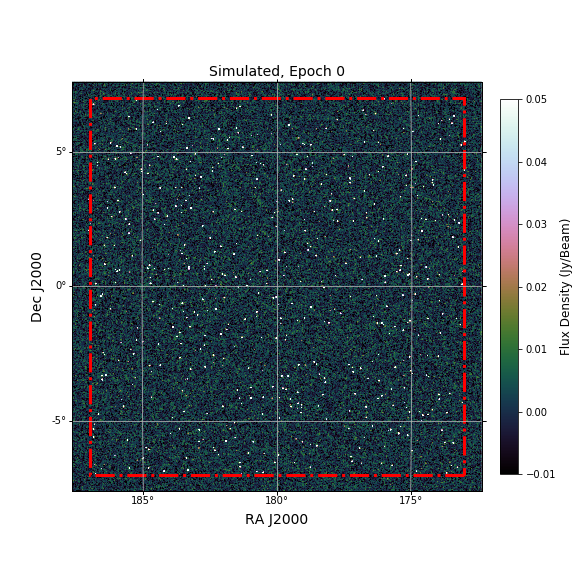}
\caption{
An image of the field-of-view for the region of interest for the first epoch of the simulated data.
The red box indicates the region of interest within which the sources were added.
}
\label{fig:simdata}
\end{figure}

The simulation does not include any positional distortions that would require \fitswarp{} in order to achieve correct source allignment.
With both the faintest source flux density and the image noise set to $5$\,mJy\,beam$^{-1}$, with 25 epochs, a (non-transient) source which is at $1\sigma$ significance in a single epoch can be detected with $5\sigma$ significance in the mean image.
The first epoch of the simulated data is shown in Figure\,\ref{fig:simdata}.

\section{Methods}
\label{sec:methods}
The data to be analyzed are modeled conceptually as a set of observations whose intersection represent an area of sky which is to be studied - the region of interest.
The region of interest consists of the following components: 
\begin{enumerate}
\item Persistent sources which may vary intrinsically or appear to vary due to propagation (or instrumental) effects,
\item Transient sources which do not appear in the reference image, but are detected in one or more individual epochs,
\item Background diffuse emission easily separable from the compact sources of interest, and
\item A noise component made up of a combination of thermal and confusion noise.
\end{enumerate}

The goal of the work-flow presented here is to separate and characterize these four components. The BAckground and Noise Estimator
(\bane{}\footnote{Part of the \href{https://github.com/PaulHancock/Aegean}{AegeanTools} library} \citep{Hancock_source_2018}) can be used to characterize the image background (3) and noise (4) components, while \aegean{} \citep{Hancock_compact_2012} can be used to extract the compact components (1+2). In a single epoch (image) the persistent and transient sources cannot be differentiated; multiple epochs (images) enable the separation of variable and transient objects by tracking the contents of the catalogues produced by \aegean{}.

The VAST pipeline \citep{banyer_vast_2012} and the TraP \citep{Swinbank_lofar_2015} treat each image separately, performing source-finding and characterization independently.
These pipelines then rely on cross-matching catalogues (source association between epochs) in order to produce the light curves for persistent sources, and to identify transient sources.
Blind source-finding, followed by cross-matching, is a difficult process because:
\begin{itemize}
\item Sources near the detection threshold will not be consistently detected in all epochs, resulting in ``drop-outs" in the light curves of such sources;
\item It is difficult to reliably associate groups of nearby sources between epochs, resulting in a ``many-to-many" problem; and
\item Close compact sources or resolved sources may not be consistently characterized between epochs, resulting in a ``split/merge" problem.
\end{itemize}

The problem of drop-outs has been addressed in the VAST/TraP pipelines by implementing a watch list of sources.
Sources that were previously found in a given region of sky, which are not found in a subsequent epoch, are added to the watch list, and from that point forward their flux densities will be measured.
This solution allows for the monitoring of transient events after their initial detection; however, it does not give any indication of the historical behavior of such events.
The problem of missing histories, is in turn resolved by reprocessing previous epochs of data to recover (limits on) the flux densities of transient sources prior to their detection.
This breaks the stream processing paradigm and is in part a motivation for the work-flow described by \tool{}.

To avoid the problems associated with blind detection followed by cross-matching, \tool{} makes use of the  priorized fitting capability of \aegean{} \citep{Hancock_source_2018}.
Priorized fitting requires an input catalogue of known sources that will be characterized: a reference catalogue.
Priorized fitting is the process of taking a reference catalogue and an image, and then for each source in the catalogue, measuring the flux density in the image at the given position, with the given source shape.
Priorized fitting differs from blind source-finding in that it is a measurement task, rather than a detection and characterization task, and thus measurements can be made for sources whose flux density may be lower than the detection threshold set in blind source-finding.
Priorized fitting guarantees that nearby and resolved sources are consistently characterized by the same number of components across all images.
Finally, the process of priorized fitting means that light curves can be generated that consist only of measurements rather than a mix of measurements and upper limits, avoiding the problems of computing statistics on masked data.
As part of its default operation, \aegean{} assigns each source a universally unique identifier \citep[UUID\footnote{https://docs.python.org/2/library/uuid.html},][]{Leach2005}.
When using the priorized fitting mode, \aegean{} provides a one-to-one mapping between the reference catalogue sources and the priorized output sources (via the UUID), so that cross matching is no longer required.

A reference catalogue can be created either from external data, or from the multiple images which are to be studied.
By combining the input images into a cube and then flattening them into a mean image, a single more sensitive image can be obtained.
\tool{} uses such a mean image to create the reference catalogue.

With this in mind it is important to note that persistent sources are those which are bright enough to be detected in the mean image, and transient sources are those which are not found in the mean image, but which can be detected in one or more of the (less sensitive) individual epochs.
It is thus quite possible that a very bright transient event can be detected in the mean image.

\subsection{Constructing a reference image/catalogue}
\label{subsec:refimage}
It is assumed that the input images all cover the same region of sky and have a 1-to-1 correspondence between pixels.
This requirement is due to the method which is used to create the image cube and mean image, and may be relaxed in the future.
With an array like the MWA this can be achieved by observing each epoch at the same local sidereal time, and then imaging with a consistent set of parameters.
In general the 1-to-1 pixel mapping can be achieved through a re-sampling of the input images using a tool such as montage\footnote{montage.ipac.caltech.edu} or {\sc SWarp} \citep{Bertin_TERAPIX_2002}.

\begin{figure}[ht]
\centering
\includegraphics[width=0.8\linewidth]{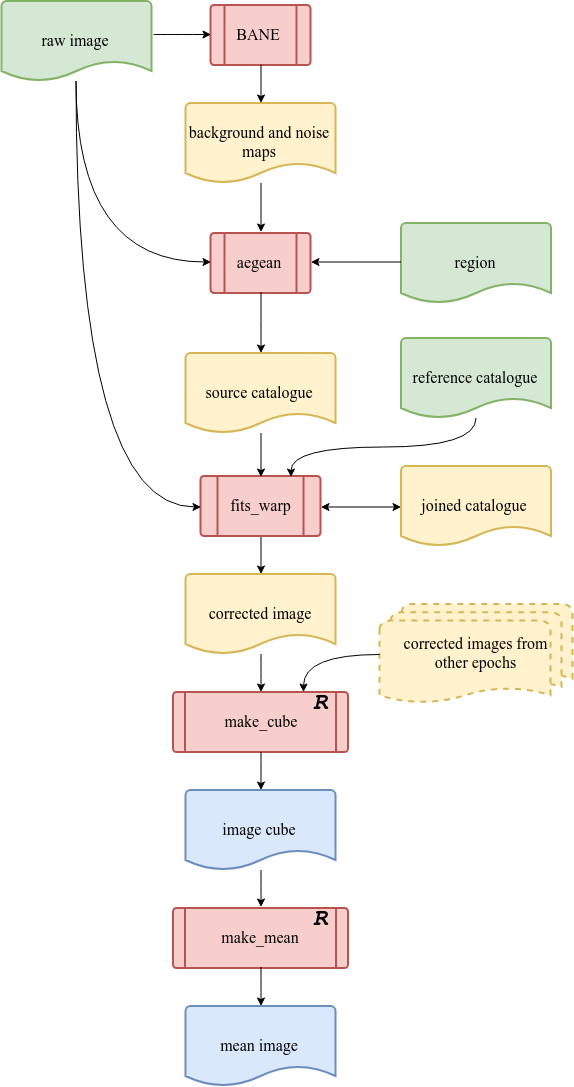}
\caption{\tool's work-flow for creating a mean image using \fitswarp{} to correct each of the single epoch images.
Green boxes represent existing input data, red boxes represent a script or process, yellow boxes represent intermediate data files, and blue boxes are data files that are common between different sections of the work-flow (\figs{} \ref{fig:priorized} and \ref{fig:transients}) .
The box with a dashed border represents the the collection of data files from parallel invocations of this work flow, each operating on a different input.
Red boxes with an {\bf R} indicate programs that are included with \tool{}, and are described in section \ref{sec:code}.
}
\label{fig:meanimage}
\end{figure}

At low radio frequencies the ionosphere can introduce a spatial warp to images, even in relatively calm conditions.
In order to undo the warping of the images, \tool{} uses \fitswarp{}\footnote{https://github.com/nhurleywalker/\href{https://github.com/nhurleywalker/fits_warp}{fits\_warp}} \citep{hurley-walker_fitswarp_2018} to correct the astrometry of each epoch individually.
The de-warping process of \fitswarp{} compares a reference catalogue to a catalogue generated from the image of interest.
The two catalogues are cross-matched, and a rubber-sheet distortion model is computed that will move sources from their observed positions to the reference positions.
This match/move cycle is repeated three times, and the final distortion model is then applied to the image pixels.
The warped pixel locations are then used to interpolate the pixel values on a regular grid as per the initial image.
The end result is an image with the same WCS and pixel coordinates as the original, but with the sources shifted to their reference positions.

The reference catalogue that is required by \fitswarp{} can either be an external reference such as the GLEAM catalogue \citep{HurleyWalker_GLEAM_2017}, or an internal reference such as a catalogue generated from an individual epoch.
An external reference catalogue will make it easier to incorporate data from other surveys, however an internal reference catalogue will still generate a consistent astrometry solution and allow for the detection of radio variables and transients.

Once the individual images have been pre-processed, \tool{} joins them together to form a cube, where the third axis of the cube is the epoch or date of observation. 
This cube is then used to create a mean image.
If the $N$ input images are thermal noise-limited (as is the case for the simulated data) then the mean image will have a lower noise by a factor of $\sqrt{N}$.
If the input images have a significant (or dominant) side-lobe or classical confusion noise component (as is the case for the observed data), then the mean image may not have a significantly lower noise.
A further consequence of confusion noise is that the image noise will be correlated between epochs.

After the creation of a reference image, \aegean{} is run in blind source-finding mode to produce a catalogue of persistent sources.
At this point the area of interest can be limited by supplying a region file in \mimas{}\footnote{Part of the \href{https://github.com/PaulHancock/Aegean}{AegeanTools} library} format.
Only sources falling within the given area of interest will be included in the catalogue of persistent sources.

\begin{figure}[h]
\centering
\includegraphics[width=0.8\linewidth]{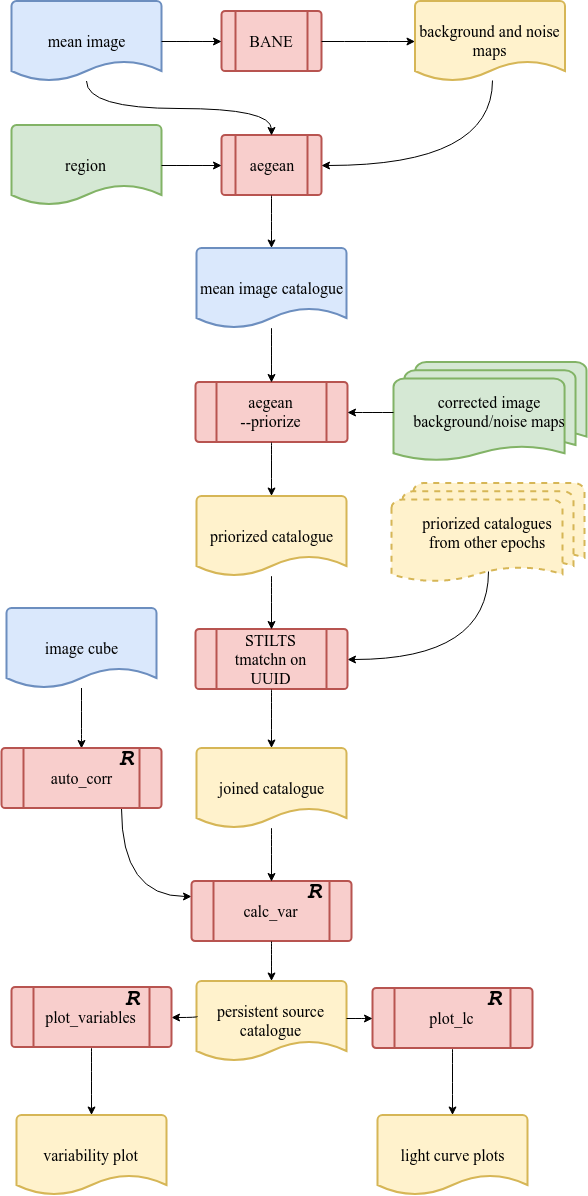}
\caption{\tool's work-flow for creating a mean image catalogue, priorized catalogues at each epoch, and final joined catalogue with variability statistics.
The colour key is as per \fig\,\ref{fig:meanimage}.
}
\label{fig:priorized}
\end{figure}

The creation of a mean image is outlined in \Fig\,\ref{fig:meanimage}.
The image shown in \Fig\,\ref{fig:skyplot} is a mean image created from the 25 epochs of the observational data.
The mean image catalogue is created from the mean image in the first steps of the work-flow outline in \Fig\,\ref{fig:priorized}.

\subsection{Light curves for persistent sources}
\label{sec:variables}

As mentioned previously, \tool{} uses priorized fitting to generate light curves for each of the persistent sources, across all epochs.
When using the priorized fitting mode, \aegean{} will copy the source shape and UUID from the reference catalogue to the output catalogue. An exact match is made across epochs using the UUID of each source, entirely avoiding cross-matching errors.
This UUID-matching is performed across all catalogues using the STILTS \citep{Taylor_topcat_2005} function \texttt{tmatchn} with \texttt{matcher=exact}.

The use of priorized fitting and matching on UUID means that every persistent source has a flux density measurement in all epochs, and there are no drop-outs, thus ensuring complete light curves. This is true even if the mean image is no more sensitive than the individual epochs (e.g. in the case of noise dominated by confusion). If the mean image \textit{is} more sensitive than the individual images, then the use of priorized fitting allows the flux densities to be measured for sources that would not have been detected when source-finding on the individual images, increasing the sensitivity of the final catalogue.

Once the light-curves have been created \tool{} characterizes them using a set of standard variability metrics: the mean flux density ($\mu$, \eq\,\ref{eq:mean}), the variance in the flux density ($\sigma$, \eq\,\ref{eq:variance}), the modulation index or coefficient of variability ($m$, \eq\,\ref{eq:m}), the de-biased modulation index ($m_d$, \eq\,\ref{eq:md}), the reduced chi-squared against a model of constant flux density ($\chi^2_{lc}$, \eq\,\ref{eq:chisquare}), and the probability of observing the given $\chi^2_{lc}$ for a non-variable source (p\_val, \eq\,\ref{eq:pval}).
Each of which are calculated as:
\begin{align}
\mu &= \frac{\sum_i S_i}{N} \label{eq:mean}\\
\sigma^2 &= \frac{\sum_i\left(\mu - S_i\right)^2}{N} \label{eq:variance}\\
m &= \frac{\sigma}{\mu} \label{eq:m}
\end{align}

\begin{align}
m_d &= \frac{1}{\mu}\sqrt{\frac{\sum_i\left(\mu - S_i\right)^2 - \sum_i \sigma_i^2}{N}} \label{eq:md}\\
\chi^2_{lc} &= \sum_i^N\frac{\left(\mu - S_i\right)^2}{\sigma_i^2} \label{eq:chisquare}\\
\mathrm{p\_val} &= 1 - \frac{\gamma\left(\frac{\chi^2_{lc}}{2},\frac{N-1}{2}\right)}{\Gamma\left(\frac{\chi^2_{lc}}{2}\right)} \label{eq:pval}
\end{align}
\noindent where $S_i$ and $\sigma_i$ are the flux density and uncertainty measured in epoch $i$, and N is the total number of epochs.
$\Gamma(x)$ is the gamma function and $\gamma(k,x)$ is the lower incomplete gamma function.
The parameter p\_val is the survival function for a $\chi^2$ distribution assuming $N-1$ degrees of freedom, and is calculated using  the \texttt{scipy.stats.chi2.sf} function \citep{Jones_scipy_2001}.
The calculation of p\_val can be adapted to account for correlated noise between images, by reducing the number of effective degrees of freedom.

As noted in Section\,\ref{subsec:refimage}, in the presence of side-lobe or confusion noise these calculations need to be modified by the covariance matrix (for $\chi^2_{lc}$), or by reducing the effective degrees of freedom (for p\_val).
\tool{} computes an effective number of degrees of freedom by sampling the image cube at random sky positions and calculating the autocorrelation function for each, and then taking the mean and variance over the sky positions.
The point at which the auto-correlation function falls to within 1 standard deviation of zero, is used to calculate the effective number of degrees of freedom, which is then used to derive p\_val.

\begin{figure}[h]
\centering
\includegraphics[width=0.9\linewidth]{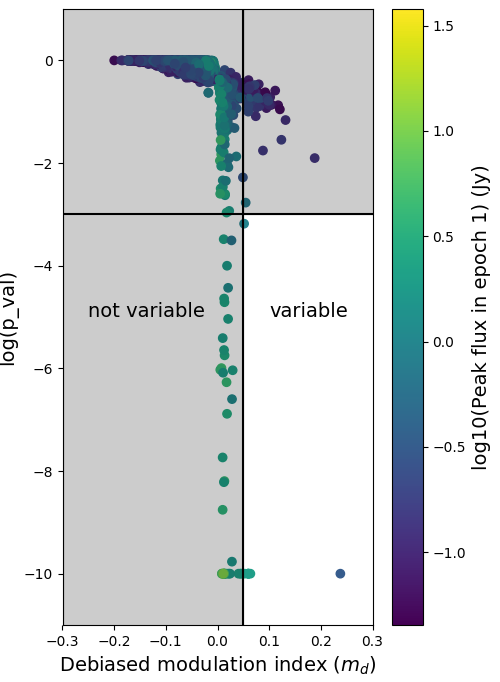}
\caption{
An example of the visualization for variable sources, from the observed MWA data.
The SNR of each candidate is indicated by the colour scale.
A source is marked as variable if the probability of being not variable is low ($\mathrm{p\_val}<1e-3$), and the magnitude of variability is high ($m_d > 0.05$), as indicated by the solid lines and unshaded area.
}
\label{fig:variablesvis}
\end{figure}

All persistent sources are reported in a combined catalogue, along with the variability statistics described above.
Additionally a plot is generated that gives a quick identification of the number of sources that are deemed to be variable.
Variable sources are selected based on the significance (p\_val), and magnitude ($m_d$) of variability.
\Fig\,\ref{fig:variablesvis} shows an example visualization of the variable/not-variable sources in the observational data.

\subsection{Transient sources}\label{sec:transients}

\begin{figure}[h]
\centering
\includegraphics[width=0.9\linewidth]{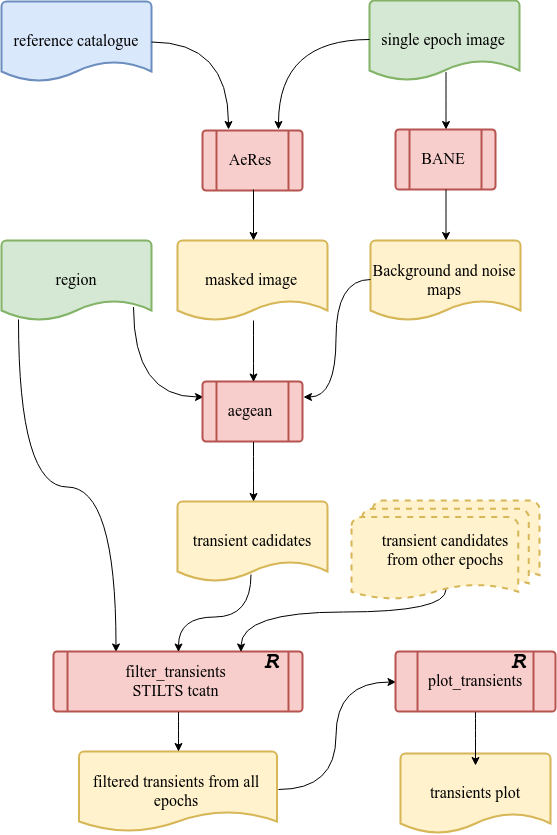}
\caption{\tool's work-flow for finding transients.
The colour key is as per \fig\,\ref{fig:meanimage}.}
\label{fig:transients}
\end{figure}

Transient sources are identified as sources that appear in an individual epoch which are not in the list of persistent sources.
In order to find such sources, \tool{} first masks all persistent sources from each epoch.
This masking is done using \aeres{}\footnote{Part of the \href{https://github.com/PaulHancock/Aegean}{AegeanTools} library}.
The masking is fairly generous - for each source, all pixels that have a model flux density above $0.1\sigma$ in the mean image are set to null value.
Note that the masking criteria is the same for each image, and thus each image will have the same pixels masked.

Once each image has been masked, \tool{} runs \aegean{} in blind source-finding mode.
Whatever sources are found in this stage are candidate transients. 
The list of candidates is contaminated by noise peaks close to modeled sources that have not been masked, but which have been brought above the detection threshold by the presence of the nearby source.
Compact components of sources which have extended emission will also not be completely filtered.
Therefore candidate transients which have a fitted position that is outside the image, outside the region of interest, or in a masked region, are removed.
The transient detection process is described in \Fig\,\ref{fig:transients}.

\begin{figure}[h!]
\centering
\includegraphics[width=0.9\linewidth]{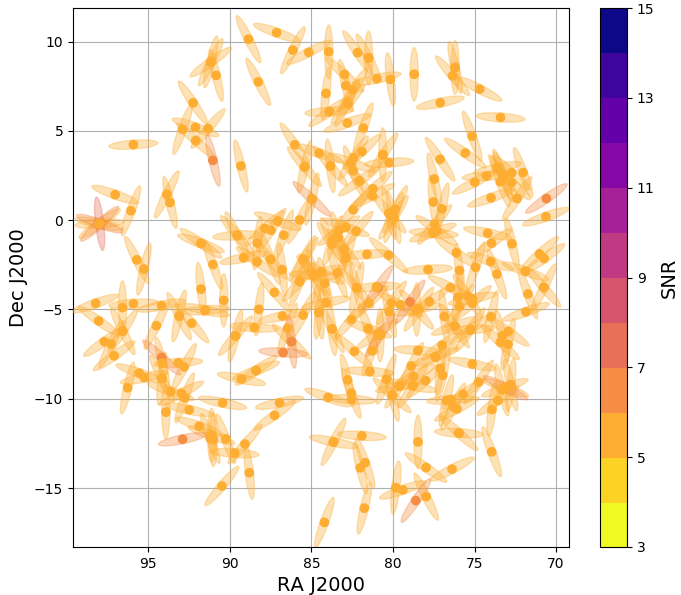}
\caption{
An example of the visualization for transient candidates using the observed data.
The SNR of each candidate is indicated by the colour scale.
Since a transient event may occur in the same location in multiple epochs, the epoch is indicated by the rotation of the ellipse.
In this example, all the transient candidates have low SNR ($\sim 5$) and are consistent with noise in individual epochs.
}
\label{fig:transientsvis}
\end{figure}

The remaining list of candidates is then presented in two formats.
The first format is a concatenation of the catalogues from each epoch, with an additional column that indicates the epoch in which the transient was detected.
The second format is an image, in which the transient locations are indicated along with the significance of the transient (using colour), and the epoch in which the transient was detected (by modifying the angle of an elliptical marker).
An example visualization for the test data is shown in \Fig\,\ref{fig:transientsvis}.

\section{Validation}
\label{sec:validation}

\tool{} was run on both the observed and simulated data sets, and the results examined to validate the software.

\subsection{Observed data}

\Fig\,\ref{fig:transientsvis} displays the transient candidates found in the observed data, all of which have SNR $\sim 5$.
The rms noise of the mean image is just 1.4$\times$ lower than that of the individual input images, due to sidelobe and classical confusion.
In the absence of confusion we would expect that the mean of 25 images would yield an rms noise 5 times smaller that the individual epochs.
Thus it is expected that persistent sources with a flux density that is below the $5\sigma$ detection threshold in the mean image will occasionally appear with a SNR$>5$ in the individual epochs and should be detected as transient candidates.
The fact that some of the transient candidates in Fig\,\ref{fig:transientsvis} appear in multiple epochs (e.g. the candidate near (97,0)) suggest that this the case.
We discount all of the transient candidates as being due to this effect and conclude that there are no true transients in the observational data.

\Fig\,\ref{fig:variablesvis} visualises the variability of the sources; the majority of sources are not variable (left part of the plot), or have low enough significance (due to their low flux densities) that their variability cannot be ascertained (upper part of the plot). All of the points which lie in the variable (lower-right) section of the plot are associated with either the Orion or Flame nebulae.
Since these nebulae are extended emission regions in the test data, we cannot trust that they correspond to true variability since the underlying source characterization (both blind and priorized) is optimized for compact sources and is known to perform poorly on regions of extended emission.
Since the Orion Nebula is a region of extended emission, which is poorly characterized by \aegean{}, this variability is discounted as being due to inconsistent characterization of the region, rather than any true variability.
We therefore class all of the variable candidates as not being real due to this effect.

The observational data contain no believable transients or variables.
This non-detection of transient or variable sources is consistent with the very low surface density reported by \citet{Bell_survey_2014} and \citet{bell_murchison_2018}.
This demonstrates the ability of \tool{} to operate on real data; however, it does not demonstrate the ability to recover real variability and identify transient events.
We therefore turn to the simulated data.

\subsection{Simulated data}

The simulated data contains 1100 sources with 500~variable sources (modulation index of 5\%), 500~non-variable sources, and 100~transient sources that are bright in only a single epoch.
The simulated data were processed using \tool{} in the same manner as the observed data, and the results are described below.

\begin{figure}[h!]
\centering
\includegraphics[width=0.9\linewidth]{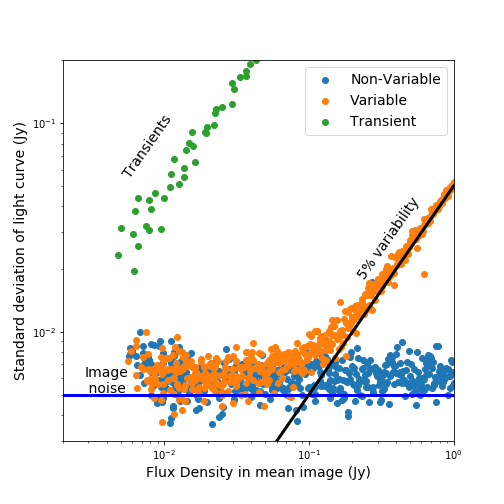}
\caption{
The variance in the light curve of a source as a function of the flux density measured in the mean image for the 1,010 sources detected.
All variable sources have a simulated modulation index of 5\%, as indicated by the black line, and the image noise is $5$\,mJy\,beam$^{-1}$, as indicated by the blue line.
The simulated source type is indicated by the color of the points.
Transient sources (green) which are bright enough to appear in the mean image are included in the persistent source catalogue and appear with high variance and low flux density in this plot.
At low flux density the variable (orange) and non-variable (blue) source population become indistinguishable, however above about 0.1\,Jy the two populations clearly diverge.
}
\label{fig:simvariables}
\end{figure}

\begin{figure}[h!]
\centering
\includegraphics[width=0.9\linewidth]{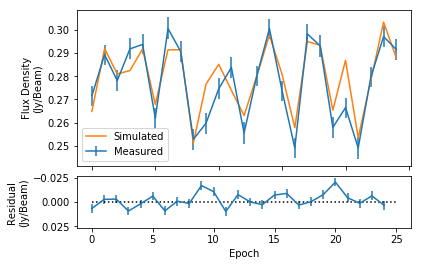}
\caption{
An example light curve for one of the simulated sources.
In orange is the simulated light curve, and in blue is the measured light curve with measurement uncertainties.
The lower panel shows the residual when the simulated light curve is subtracted from the measured light curve.
Subtracting the input light curve from the measured light curve reduces the $\chi^2_{lc}$ from 347 to 71, so that the residual light curve would have a $\mathrm{p\_val}$ and $m_d$ that put it squarely in the not variable category.
}
\label{fig:singlesimlc}
\end{figure}

Figure\,\ref{fig:simvariables} shows the measured variance in the light curve of all the persistent sources, as well as their flux densities as measured in the mean image.
There are three source populations visible: sources that have low variance, comparable to the image noise; sources with excess variance that is consistent with the 5\% input variability; and sources with extreme variance well above the 5\% level.
As indicated by the coloring of the points in \Fig\,\ref{fig:simvariables}, the sources with excess variance are either transients or variables, while the sources with low variance are either non-variable sources, or faint variable sources.
The input and output light curves for a variable source of intermediate brightness are shown in \Fig\,\ref{fig:singlesimlc}: \tool{} provides correct and believable light curves.

\begin{figure}[h!]
\centering
\includegraphics[width=0.9\linewidth]{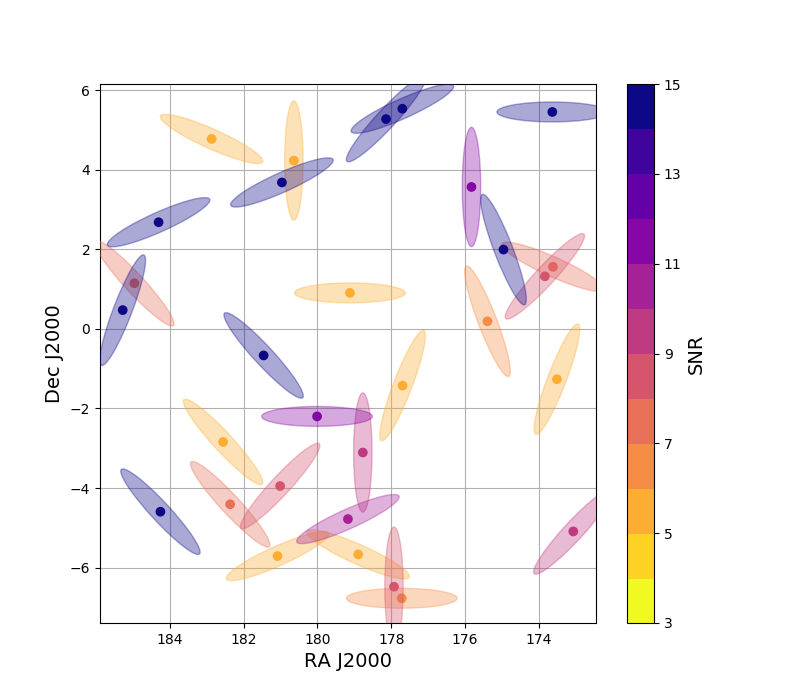}
\caption{
An example of the visualization for transient candidates using the simulated data.
The SNR of each candidate is indicated by the colour scale.
Since a transient event may occur in the same location in multiple epochs, the epoch is indicated by the rotation of the ellipse.
All of the candidates match with a known input transient source in the simulated data.
}
\label{fig:simtransients}
\end{figure}

When generating the simulation, we know which sources are variable, and which are transient. However, \tool{} distinguishes between persistent and non-persistent sources based on whether they appear in the mean image.
The persistent sources yield a catalogue such as the one that was used to generate \Fig\,\ref{fig:variablesvis}, whilst the non-persistent sources are stored in a transients candidate list which is used to generate \Fig\,\ref{fig:simtransients}.
Since the simulated transients are only non-zero flux density in a single epoch, there are a number of expected outcomes for these sources depending on their brightness:
\begin{enumerate}
    \item Bright sources will appear in the mean image at $1/25^{\mathrm{th}}$ of their peak flux density, be classified as persistent, and appear in the variable source list with a very large modulation index, and their light curves will be characteristic of transients;
    \item Intermediate-brightness sources will not appear in the mean image, but will be detected in the masked version of the single epoch image, and be included in the list of transient candidates that is produced by \tool{}; and
    \item Faint transient sources will be too faint to detect even in the single epoch images and will not be picked up by \tool{} at all.
\end{enumerate}

Additionally, a source of any type may be missed by \tool{} if it lies close to another source in the mean image.
When two sources are very close together \aegean{} will characterize them as a single component, and \tool{} will then generate a single light curve.
In the validation analysis that is presented here, the cross-matching of input to output sources assumes a one-to-one mapping, and thus two very close input sources will result in only a single match to the output source.
To account for this behaviour in our validation test, we manually inspect each of the input sources which are not found in the mean image and classify them as either being below the detection threshold, or confused with another source.

Of the 1100 simulated sources \tool{} found 970 persistent sources, and 24 transient candidates.
Of the simulated 500 variable sources, 480 were detected as persistent sources, 9 were faint sources not detected at all, and 11 sources were too close to a brighter source to be detected individually.
Of the simulated 500 non-variable sources, 490 were detected as persistent sources, 7 were faint sources not detected at all, and 3 were confused with brighter sources in the mean image.
Of the 100 simulated transient sources, 40 were bright enough to be detected in the mean image and are seen as high variance variable sources in \fig\,\ref{fig:simvariables}, 24 intermediate brightness sources were included in the list of transient candidates, and 31 sources were too faint to be detected in either the mean or the single epoch images, while 2 sources were confused with brighter sources in the mean image, and 3 were not detected even though they were above the $5\sigma$ detection limit in the single epoch image.
\tab\,\ref{tab:simrecovered} summarizes the success rate of detecting sources of each type.

\begin{table*}[hbt]
    \centering
    \begin{tabular}{c|c|c|c|c|c}
        \hline
         Type        & Simulated & Found in   & Below $5\sigma$ & Candidate  & Missed \\
                     &           & mean image & in mean image & transients & \\
        \hline
        Not variable & 500      & 490         & - & -  & 10 (2\%) \\
        Variable     & 500      & 480         & - & -  & 20 (4\%) \\
        Transient    & 100      & 40          & 31 & 24 & 5 (7\%$^\dagger$) \\
        \hline
    \end{tabular}
    \caption{
    The success rate of recovering different simulated source types.
    Sources were either detected in the mean image or in the transients candidate list.
    $^\dagger$Note: Excluding the 31 sources that were too faint to be detected in either the mean image or single epochs.
    }
    \label{tab:simrecovered}
\end{table*}

The simulated data show clearly that \tool{} is able to extract believable light curves which resemble the 'true' light curves, and is able to capture transient events either through the transients candidate list or via outliers in the persistent source catalogue.

\section{Code}
\label{sec:code}
\tool's full processing work-flow is encapsulated in a Makefile and related Python scripts.
The code can be downloaded from GitHub\footnote{\github\href{https://github.com/PaulHancock/Robbie}{Robbie}}.
The version described in this document has commit hash ff839e7. %COMMIT HASH

The following Python scripts are included:
\begin{itemize}
\item \texttt{calc\_var.py}
\item \texttt{auto\_corr.py}
\item \texttt{make\_cube.py}
\item \texttt{make\_mean.py}
\item \texttt{plot\_lc.py}
\item \texttt{plot\_variables.py}
\item \texttt{filter\_transients.py}
\item \texttt{plot\_transients.py}
\end{itemize}

The included scripts are described briefly below.

\texttt{calc\_var.py}:
calculate statistics for a joined table of sources.
The table is assumed to be joined horizontally without any blank entries, and with each epoch being distinguished via a suffix in the column names.
The stats that are calculated are: the mean peak flux density, the modulation index, the de-biased modulation index, the reduced chi-squared against a model of constant flux density, the p-value corresponding to the reduced chi-squared.
By default the number of degrees of freedom are set to one less than the number of epochs, however this can be changed via a command line flag.
The effective number of degrees of freedom can be calculated using \texttt{auto\_corr.py}.
The new columns are appended to the existing table, and written to the output file.

\texttt{auto\_corr.py}: analyze an image cube and determine an effective number of degrees of freedom by calculating the average autocorrelation across number of sky positions.

\texttt{make\_cube.py}:
take an list of input files, read a 2D image from each, and create a 3D cube which is then written to disk.
The fits header of the first file is copied to the output file.
It is required that the input images all have the same pixel dimensions and pixel to sky coordinate mapping. 

\texttt{make\_mean.py}:
flatten a 3D image cube into a 2D mean image.
The flattening operation is to take the (un-weighted) mean of the pixels across epochs.

\begin{figure}
\centering
\includegraphics[width=0.9\linewidth]{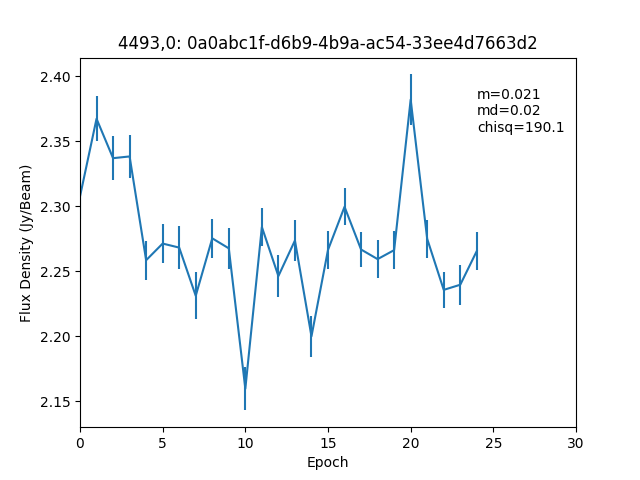}
\caption{An example light curve generated by \tool{}.
The figure title is the catalogue row number followed by the UUID of the source.}
\label{fig:lightcurve}
\end{figure}

\texttt{plot\_lc.py}:
for a given input catalogue of persistent sources, this script will create light curves.
The plot file names are the UUID of each source, and each plot will be annotated with: catalogue row number, UUID, modulation index, de-biased modulation index, and $\chi^2_{lc}$.
An example plot from the observed data ise shown in \fig\,\ref{fig:lightcurve}.

\texttt{plot\_variables.py}:
given an input catalogue of persistent sources, this script will create a variability visualization as per \Fig\,\ref{fig:variablesvis}.

\texttt{filter\_transients.py}:
take a list of transient candidates, and remove those which are considered to be obviously spurious.
Sources which have a fitted position that is in a masked region of the corresponding image, are removed as being false detections.
Sources with a position that is outside of the specified region of interest are also considered to be spurious.

\texttt{plot\_transients.py}:
given an input catalogue of transient candidates, this script will create a transients visualization as per \Fig\,\ref{fig:transientsvis}.

\section{Summary}
We have presented a description of \tool: a batch processing work-flow for the detection of radio transients and variables.
\tool{} is designed to be a modular work-flow that can be adjusted to accommodate a variety of radio image data, whilst maintaining a consistent methodology.
We have used both real and simulated data to demonstrate and validate the capabilities of \tool{}.
In order to increase the portability of \tool{} and facilitate its use on HPC scale facilities, \tool{} will need to be ported from using {\sc Make} as its work-flow manager to a package like CWL, which can better integrate with job schedulers and Docker containers.
Such changes will need to be incorporated before \tool{} can begin to process the volumes of data at the rate required for the Square Kilometer Array (SKA).

We encourage interested parties to contribute to this open source project by submitting feature enhancement and pull requests via the GitHub page at: \href{\github{}Robbie}{\github{}Robbie}.

\section*{Acknowledgements}
\subsection*{People}
We thank the two referees for their feedback which has greatly improved this publication.
\subsection*{Software}
% python modules
We acknowledge the work and support of the developers of the following following Python packages: Astropy \citet{TheAstropyCollaboration2013,TheAstropyCollaboration2018}, Numpy \citep{vanderWalt_numpy_2011}, Scipy \citep{Jones_scipy_2001}, Pandas \citep{mckinney_pandas_2010}.
% prereqs
\tool{} relies upon the following software: AegeanTools \citep{Hancock_source_2018}, TOPCAT \citep{Taylor_topcat_2005}, and \fitswarp{} \citep{hurley-walker_fitswarp_2018}.
% visualisation tools
Development of \tool{} made extensive use of DS9\footnote{\href{ds9.si.edu}{http://ds9.si.edu/site/Home.html}} and TOPCAT for visualization. 
% software
This research made use of Astropy, a community-developed core Python package for Astronomy \citep{TheAstropyCollaboration2013,TheAstropyCollaboration2018}.

\subsection*{Facilities}
% computing resources
This work was supported by resources provided by the Pawsey Supercomputing Centre with funding from the Australian Government and the Government of Western Australia.
% MWA
This scientific work makes use of the Murchison Radio-astronomy Observatory, operated by CSIRO. We acknowledge the Wajarri Yamatji people as the traditional owners of the Observatory site. 

\bibliographystyle{model2-names}
\bibliography{mybibfile}
\end{document}